\documentclass[journal,10pt]{IEEEtran}
\usepackage{graphics}
\usepackage{graphicx}
\usepackage{latexsym}
\usepackage{amssymb}
\usepackage{amsmath}
\usepackage{multirow}
\usepackage{epstopdf}
\usepackage{CJKutf8}
\usepackage{amsfonts}
\usepackage{setspace}
\usepackage{tabularx,ragged2e,booktabs,caption}
\usepackage{cite}
\usepackage{tabulary}
\usepackage{color}
\usepackage{epsfig}
\usepackage{subfigure}
\usepackage{subfig}
\usepackage{lipsum}
\usepackage{makecell}
\usepackage{glossaries}
\usepackage[utf8]{inputenc}
\newtheorem{prop}{\bf{Proposition}}

\begin{document}
\setlength{\abovedisplayskip}{1pt}
\setlength{\belowdisplayskip}{1pt}
\title{Offloading of Users in NOMA-HetNet Using Repulsive Point Process}


%
%
\author{Pragya Swami,  Vimal Bhatia, \emph{Senior Member, IEEE}, Satyanarayana Vuppala, \emph{Member, IEEE},  \\ and Tharmalingam Ratnarajah, \emph{Senior Member, IEEE}
	\vspace{-2ex}
	\thanks{This work is supported by the MeitY, Indian Institute of Technology Indore, Indore, India, and University of Edinburgh, Edinburgh, UK.}
	\thanks{P. Swami and V. Bhatia are with Indian Institute of Technology Indore, Indore, India; S.~Vuppala is with University of Luxembourg, Luxembourg; and T. Ratnarajah is with the Institute for Digital Communications, University of Edinburgh, Edinburgh, UK.}
}

{}
\maketitle

\begin{abstract}
Ever increasing number of cellular users and their high data requirements, necessitates need for improvement in the present heterogeneous cellular networks (HetNet). Carrier sensing prevents base stations within a certain range of the transmitter from transmitting and hence aids in reducing the interference. Non-orthogonal multiple access (NOMA) has proven its superiority for the 5th generation (5G) networks. This work proposes a mathematical model for an improved HetNet with macro base station (MBS) and femto base station (FBS) tier. The FBS tier is equipped to support NOMA and carrier sensing for its transmissions. Offloading is performed for load balancing in HetNet where the macro users (MU) from congested MBS tier are offloaded to the FBS tier. The FBS tier pairs the offloaded MU (OMU) with an appropriate pairing user (PU) to perform NOMA. The performance of the OMU is studied under different channel conditions with respect to the available PU at the FBS and some useful observations are drawn. A decrease in outage probability by $74.04\%$ for cell center user (CCU) and $48.65\%$ for cell edge user (CEU) is observed for low density FBS. The outage probability decreases by $99.60\%$, for both the CCU and CEU, for high density FBS using the proposed carrier sensing in NOMA. The results are validated using simulations.
\end{abstract}
\begin{IEEEkeywords}
	Non-orthogonal multiple access, stochastic geometry, repulsive point process, heterogeneous cellular network, outage probability.
\end{IEEEkeywords}
\IEEEpeerreviewmaketitle

\section{Introduction}
Femto base station (FBS) deployment in the existing cellular network is one of the most viable solution to meet the intense consumer demands for mobile data while catering to ever increasing number of cellular users. The resulting network of macro base station (MBS) and FBS, termed as heterogeneous cellular networks (HetNet), provides a cost-effective expansion to existing cellular wireless networks. To increase this capacity further, non-orthogonal multiple access (NOMA), now included in the Release 15 of 3rd generation partnership project (3GPP), has gained wide interest recently as an enabling technique for 5th generation (5G) mobile networks and
beyond. NOMA has proven to provide better spectral efficiency \cite{random}, \cite{saito} as compared to orthogonal multiple access (OMA) adopted by 4G mobile communication systems standardized by 3GPP such as Long Term Evolution (LTE) \cite{LTE} and LTE-Advanced \cite{LTEA}.   

In a HetNet, the FBS acts as an offloading spot and helps in load balancing by serving some of the macro users (MU), called as offloaded user (OU) at the FBS, from the congested MBS tier \cite{offloadingsarabjit}, \cite{VTC}. The FBSs are deployed opportunistically or randomly making conventional frequency planning strategies very difficult (and redundant) in a two-tier network \cite{pastpresentfuture}, \cite{interferencesystem}. The lack of coordination between FBSs leads to randomized co-tier interference which results in performance degradation. This interference can be mitigated by using a medium access channel (MAC) protocol, as one of the possible solution, involving carrier sensing in the FBS tier for interference management such that the transmitting FBS does not end up using the channel that is already occupied by other FBS. Carrier sensing forbids the FBS contending for the same channel to transmit simultaneously. 

To do this, each FBS senses the channel and transmits only if the channel is not occupied by any other contender. The distance within which carrier sensing is performed is called as contention radius (CR) and the FBSs within CR are called as contenders. Clearly, one of the contender wins and accesses the spectrum. Hence, we can say that this carrier sensing creates an exclusion region (equals to CR) around a FBS within which no other FBSs are allowed to transmit. The exclusion region around a FBS can be visualized as an existence of a minimum distance, equal to CR, between the FBSs. This makes the FBS's positions correlated with other FBSs, since, it is required to maintain a minimum distance between FBSs. The formation of exclusion region around FBSs can be modeled spatially using repulsive point processes (RPP) for e.g. hard core point process (HCPP) with a hard core parameter (HCP). While modeling the base stations using RPP, the HCP physically equals the CR within which the base stations contend for spectrum access. The Poisson point process (PPP) model assumes no correlation amongst the nodes' position thereby rendering PPP assumptions inaccurate for modeling the active transmitters that coordinates for spectrum access using carrier sensing \cite{stochasticsurvey}. The inaccuracy of PPP to model location of base stations (BS) for different tiers of HetNet is demonstrated in \cite{PPP1,PPP2,pinto,andrewsPPPproblem}. To capture the characteristics of cellular networks using carrier sensing or MAC protocol, point processes for e.g., RPP where distances among BS are fixed have proved to be more accurate than the PPP assumptions \cite{stochasticsurvey}, \cite{elsawy}. Thus, in this work we consider and analyze RPP for modeling the FBS tier.

Bertil Mat{\'e}rn proposed three approaches to construct a RPP from parent PPP leading to the formation of Type I, Type II, and Type III Mat{\'e}rn Hard core point process (MHCPP). Here, primary points are used to refer to the points in the parent PPP while secondary points are used to refer to the points of the constructed MHCPP. Type I MHCPP deletes all the primary points from the parent PPP that coexists within a distance less than the HCP. The construction of Type II MHCPP requires every point to be associated with a time mark and deletes the primary points coexisting within a distance smaller than the HCP, provided it also has a lowest time mark. However, this method leads to underestimating the intensity of simultaneously active transmitters \cite{elsawy}, \cite{INRIA}. Type III  MHCPP removes this flaw by following similar procedure as that for Type II, however, the primary point is deleted only if it coexist within a distance less than the HCP from another secondary point with a lower time mark. The aggregate interference for cognitive radio network under MHCPP is characterized in \cite{M1}, \cite{M2}, which is later approximated as a PPP model assuming fading and shadowing effects.

The MBS are generally studied as serving users with similar requirements, hence it distributes its power equally amongst them. On the other hand, for the deployed FBSs the range of user requirements varies from ultra high definition video transmissions to low power sensors in an Internet of Things setup \cite{IoT}, \cite{NOMAIoT}. It may also be required to fulfill such varied requirements simultaneously. Hence, in this work, we use power splitting amongst the users appropriately using NOMA in FBSs to support the offloaded users. NOMA uses superimposition of users' signal with different channel conditions in power domain unlike OMA \cite{LTE} which uses orthogonality in frequency, time or code to serve multiple users. In the literature, most system models that employ NOMA generally do not account for practical system characteristics such as the minimum distance constraint between the base stations. For instance, \cite{random}, \cite{underlay} consider PPP distributed base stations (BS) for their analysis without employing any MAC protocol. Therefore, advanced system models with more realistic approach should be analyzed. Since, power domain NOMA involves superimposing signals of users with different channel conditions in power domain; to identify and differentiate the femto users (FU) with different channel conditions we assume two types of users namely cell center user (CCU) and cell edge user (CEU). CCU being close to the FBS as compared to the CEU has better channel condition than CEU. The difference in channel conditions leads to different performance at the two types of users, hence, user fairness needs to be maintained which is studied in the context of NOMA in \cite{userfairnessMIMONOMA}. CEU performance and user fairness is also studied in \cite{VTC, CEU2,fairnesssystem,fairnesssystem2}.

\subsection{Motivation and Contribution}
Motivated by the need of a carrier sensing in the current HetNet to model the randomized FBS tier, for interference management, the usefulness of the concept of offloading for load balancing in the congested HetNet, and the advances of NOMA to meet the requirements of 5G and beyond services, we propose a framework of an offloading model in HetNet using NOMA at FBS tier, referred as NOMA-HetNet. The FBS tier also uses carrier sensing to manage the interference caused by their dense and random deployments. The carrier sensing in FBS tier with NOMA is modeled using an RPP\footnote{The terms RPP and carrier sensing are used interchangeably in the paper}, as explained in Section~\ref{sec:retaining} since, the PPP assumptions renders inaccurate results for modeling correlated points that occur as a result of carrier sensing. Towards this end, to the best of our knowledge, there exists no literature which studies and analyzes the impact of RPP with NOMA in the offloading environment of HetNets.

The key contributions of this work are listed below
\begin{itemize}
\item An analytical framework is designed for a HetNet where the FBS tier is equipped with carrier sensing for interference management and NOMA for power splitting. 
  
\item To model the carrier sensing amongst the FBSs we use RPP modeling. A retaining model for the RPP is explained in Section~\ref{sec:retaining} to decide the density (or number) of active FBSs based on the carrier sensing used. 

\item Offloading is performed for load balancing by handing some users from the congested MBS to the FBS tier. Since, FBS uses NOMA to support the offloaded users, it pairs the incoming offloaded macro user (OMU) with an available pairing user (PU). Also with NOMA, it becomes important to know whether the OMU is a CCU or CEU with respect to the available OMU. Hence, the concept of NOMA compatibility is discussed and the impact of offloading is analyzed.

\item A comparative study between the HetNet using PPP and RPP for modeling FBS tier with NOMA is performed. To make the study broader we also include the comparison of the proposed model with that of HetNet using OMA technique and modeling based on PPP.
\end{itemize}
Rest of the paper is organized as follows. The general system model is given in Section~\ref{sec:GSM} and the retaining model is discussed in Section~\ref{sec:retaining}. Section~\ref{sec:analytical} derives some useful expressions of outage probabilities for the proposed model using RPP. Numerical results are discussed in Section~\ref{sec:results}. Finally, the work is concluded in Section~\ref{sec:conclusion}.
\section{General System Model} \label{sec:GSM}
\begin{figure}
	\center
	\includegraphics[height=5.5cm,width=9cm]{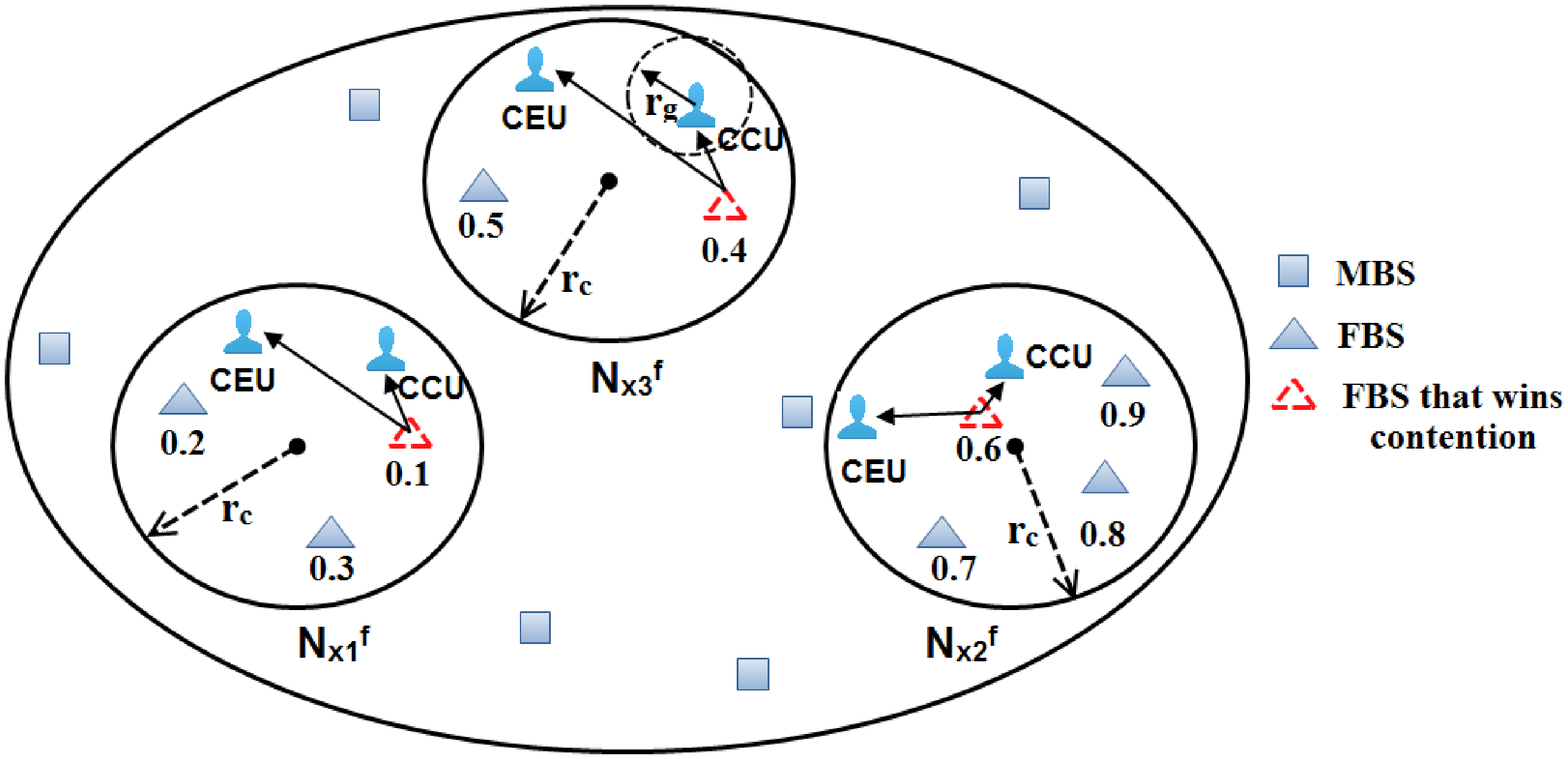}\\
	\caption{ System Model }\label{fig:system}
\end{figure}

A HetNet comprising of MBS underlaid with FBS is considered in the analysis where the FBS tier supports offloaded users from congested MBS tier for load balancing. FBS tier employs NOMA (hence also referred as FBS-NOMA) and carrier sensing for transmission. We assume that $\Omega_{m}$ denotes the PPP distributed nodes for MBS tier with intensity $\lambda_{m}$ and $\Omega_{u}$ denotes the PPP distributed nodes for users with intensity $\lambda_{u}$. $\Omega_{f}$ denotes the marked PPP (refer Fig.~\ref{fig:system}) with intensity $\lambda_{f}$, i.e, $\Omega_{f} \in {(x_i^f, p_i^f); i = 1, 2, 3, \ldots }$, where $x_i^f$ denotes the position of $i^{th}$ FBS with associated time mark, $p_i^f$, distributed uniformly in the range $[0, 1]$. The marked PPP distribution of FBS tier underlaid with MBS tier is shown in Fig.~\ref{fig:system}. The FBS that are retained using carrier sensing is explained in detail in Section~\ref{sec:retaining}. $\Omega_{f}^R$ denotes the set constructed using carrier sensing at FBS tier. It should be noted that the MBS tier does not perform carrier sensing. Assuming $t \in \{m,f\}$ denoting MBS, FBS tier, respectively, the transmit power of $t^{th}$ tier is denoted by $P_t$ and the target rate of a typical user for both the tiers is represented by $R$. $\mathcal{Y}_t$ denotes the coverage range of the BS of $t^{th}$ tier. Bounded path loss model is considered as $L(r_t)=\frac{1}{1+r_t^{\alpha_t}}$ which ensures that path loss is always smaller than one even for small distances \cite{shotnoise}, where $r_t$ is the distance between the typical user and the associated BS of $t^{th}$ tier and $\alpha_t$ is the path loss exponent of $t^{th}$ tier. Hence, the total channel gain is given by $\vert h \vert^2=\vert \hat{h} \vert^2 L(r_t)$, where $\vert \hat{h}\vert^2$ is Rayleigh distributed. The overall system transmission bandwidth is assumed to be 1 Hz. We assume a guard zone of radius $r_g >1$ around a receiver as shown in Fig.~\ref{fig:system}. The interference at a receiver is calculated beyond this guard zone. 

{\bf{\textit{Note}:}} Throughout the paper $\vert \hat{h} \vert^2$ implies Rayleigh distribution, $\vert \tilde{h} \vert^2$ will denote unordered channel gains and $\vert {h} \vert^2$ will imply ordered channel gain.

\section{Retaining Model for FBS Tier}\label{sec:retaining}
In this section, we derive the density of active FBS retained under the applied carrier sensing. The process of finding the active FBSs using carrier sensing is termed as thinning process. The marked PPP model provides the baseline model (or parent model) for the distribution of all FBSs while the subset of FBSs that succeed to access the spectrum will be modeled using RPP. The parent model distribution consists of uniformly marked PPP as shown in Fig.~\ref{fig:system} and the contention radius is denoted by $r_c$. To carry out the contention, we first find the neighborhood set of a generic FBS, $x_i^f$, contending for the channel. The neighborhood set of generic FBS is denoted by $\mathcal{N}_{x_i^f}$. The notion of received signal strength is used to decide the neighborhood set of a generic FBS, mathematically written as $N_{x_i^f} = \{(x_i^f, p_i^f) \in \Omega_{f} | \gamma (x_i^f, x_j^f) > T_B\}_{i\ne j}$ where $\gamma (a,b)$ denotes the received SNR at node $a$ from node $b$. This implies that $N_{x_i^f}$ is the set of neighbor FBSs such that the received signal-to-noise ratio (SNR) at the generic FBS at $x_i^f$ is greater than the BS-sensing threshold, $T_B$. The criterion for selecting the FBS amongst all the FBS in the neighborhood is based on their time marks. The FBS that qualifies to transmit (or alternatively we may refer to the FBS which is retained) is determined by the lowest time mark amongst its neighborhood set as can be seen from Fig~\ref{fig:system}. For the three neighborhood set shown in Fig~\ref{fig:system}, namely $N_{x_1^f}, N_{x_2^f}$, and $ N_{x_3^f}$, the FBS that wins the contention carries the lowest marks amongst all the other neighbors. This method is similar to the general carrier sensing multiple access (CSMA) protocol \cite{CSMAfollowsRPP1}, \cite{CSMAfollowsRPP2}. Following this procedure, we find the retaining probability of FBS under the above conditions. 

In the first step, the contention radius, $r_c$, is calculated. By bounding the observation to the region $B_{x_i^f}(r_c)$, gives us all the FBS inside the radius of $r_c$ centered at $x_i^f$. The contention radius is taken to be sufficiently large such that the probability of an FBS in the neighborhood of $x_i^f$ lying beyond $r_c$ is negligible. Mathematically, we write it as
\begin{equation}\label{eq:contentioncondition}
\mathcal{P}\left\{\rho_f \vert \hat{h}_(i,j) \vert^{2} L(r_j)> T_B\vert r_j > r_{c}\right\}\leq\epsilon,
\end{equation}
where $\rho_f=P_f/\sigma_f^2$ denotes the transmit SNR of FBS tier, $\sigma_f^2$ is the noise variance, $r_j$ is the distance of $j^{th}$ BS in disc $B_{x_i^f}(r_c)$ to BS $x_i^f$.
By rearranging (\ref{eq:contentioncondition}), we calculate the contention radius as,
\begin{equation}\label{eq:contentionradius}
r_{c}=\left(\frac{\rho_f \vert \hat{h}_j \vert^2 L(r_j)}{T_B} {F}_{X}^{-1}(\epsilon)\right)^{1/\alpha_f},
\end{equation}
where $X=\vert \hat{h}_j \vert^2$, ${F}_{X}^{-1}(\epsilon)$ represents the inverse of the cumulative distribution function (CDF) of the fading distribution evaluated at infinitesimal $\epsilon$. 
The neighborhood success probability (NSP) is defined as the probability that an FBS $x_j^f$ qualifies the minimum signal strength of $T_B$ at $x_i^f$ and become its neighbor. Amongst the neighboring FBS in $N_{x_i^f}$, the FBS with the lowest time mark wins the contention and is allowed to access the channel. The NSP is calculated using (\ref{eq:contentionradius}) as
\begin{equation}\label{eq:NSP1}
\mathcal{P}_{s}=\mathcal{P}\left\{\rho_f X L(r_j)\geq T_B \vert x_{j}^f\in \mathcal{B}_{x_{i}^f}(r_{c})\right\}.
\end{equation}
With the assumption of Rayleigh faded channel and bounded path loss model we write (\ref{eq:NSP1}) as
\begin{multline} \label{eq:NSP2}
\mathcal{P}_{s}=\int_{0}^{1} f(r_j) \left(1-F_{X}\left(\frac{T_B r_j^{\alpha_f} }{\rho_f}\right)\right)\mathrm{d}r_j + \\ \int_{1}^{r_c} f(r_j)\left(1-F_{X}\left(\frac{T_B r_j^{\alpha_f} }{\rho_f}\right)\right)\mathrm{d}r_j, 
\end{multline}
where $f(r_j)=2 r_j/ r_c^2$. Solving (\ref{eq:NSP2}) we get the NSP as
\begin{multline}\label{eq:NSP}
\mathcal{P}_{s}=\frac{1}{r_c^2} e^{-\frac{T_B}{\rho_f}}+ \frac{ 2 ({T_B}/{\rho_f})^{-2/\alpha_f } \Gamma \left(\frac{2}{\alpha_f },{T_B}/{\rho_f}\right)}{\alpha_f  r_c^2}-\\
\frac{2 \left(({T_B}/{\rho_f}) r_c^{\alpha_f }\right)^{-2/\alpha_f } \Gamma \left(\frac{2}{\alpha_f },{T_B}/{\rho_f} r_c^{\alpha_f
	}\right)}{\alpha_f},
\end{multline}
where $\Gamma(a, x)= \int_{x}^{\infty} e^{-t} t^{a-1}  dt$ is the incomplete gamma function. From \cite{elsawy}, we can directly write the retaining probability for CSMA protocol as $\mathcal{P}_{R} = \frac {1-e^{N_e P_{s}}}{N_e P_{s}} $, with the expected number of FBSs in the disc of radius $r_c$ around $x_i^f$, i.e. in $B_{x_i^f}(r_c)$, as $N_e = \pi \lambda_f r_c^2 $.

\textbf{Remark 1:} The NSP in (\ref{eq:NSP}) shows a vital dependence on the selection of $r_c$.  As we increase the $r_c$, the probability of a FBS lying in the neighborhood of generic FBS decreases since the received SNR decreases with increase in distance between the base stations. Also, the $r_c$ needs to be selected sufficiently large such that NSP beyond $r_c$ is negligible. Hence, $r_c$ needs to be selected appropriately. Then, the intensity of active number of transmitting FBS using carrier sensing is given by $\lambda_{f}^R = \lambda_{f} \mathcal{P}_{R}$.

\subsection{Signal to Interference and Noise Ratio at Typical Macro User}\label{sec:SINRMU}
Given the signal intended for a typical macro user (MU) as $x_m$, the signal transmitted by the MBS can be written as $X_{m,tx}=\sqrt{P_m} x_m$ and the received signal can be written as $X_{m,rx}=\sqrt{P_m} x_m \tilde{h}_m +n_m$, where $n_m$ denotes channel noise at MBS tier. The useful signal power, noise and/or interference power for a given signal $X$ can be easily calculated using $P=\mathbb{E}\left[XX^{*}\right]$, where $\mathbb{E}[.]$ denotes the statistical expectation. Hence, the signal to interference and noise ratio (SINR) at a typical MU, normalized by noise variance, can be written as
\begin{equation}\label{eq:SINRMBS}
\text{SINR}_m=\frac {P_m \rho_m^r \vert \tilde{h}_m \vert^2}{\rho_{f}  \mathcal{I}_{f}+1},
\end{equation}
where $\rho_m^r=\mathbb{E}\left[x_m^2\right]/ \sigma_m^2$ and $\rho_{f}=P_{f}/ \sigma_f^2$ denotes the receiving transmit SNR of MBS and the transmit SNR of FBS tier, respectively, and $\sigma_m^2$ and $\sigma_f^2$ represents the noise variance of MBS and FBS tier, respectively. $\mathcal{I}_{f}= \sum _{v\in \Omega_{m}^R/\{0\}} \vert \tilde{h}_v \vert ^2$, where $\vert \tilde{h}_{v}\vert^2$ denotes the total channel gain from $v^{th}$ FBS (except at origin) to typical MU at the origin. It should be noted that we have assumed orthogonality in the MBS tier hence, the co-tier interference is neglected for the analysis of MBS tier. However, for the FBS tier, we consider both the cross-tier interference as well as the co-tier interference.

\subsection{SINR at Typical Femto User (with NOMA)}

Let us assume that M femto users (FUs) are being served by an FBS. The channel gains of the FUs are ordered as $\vert{h_1}{\vert^2} \leq \cdots \leq \vert{h_{M_f}}{\vert^2}$ and their respective power allocation factors are ordered as as ${a_1} \geq \cdots \geq {a_{M_f}}$. Given $x_i$ as the intended signal for $i^{th}$ FU such that $\mathbb{E}[x_i^2]$ is assumed to be equal $\forall{i} \in{(1,2,\cdots,{M_f})}$. The signal transmitted by the FBS is given by $X_{f}=\sum_{i=1}^{M_f} x_i \sqrt{a_i P_f}$. Hence, the signal received at $k^{th}$ typical FU (which can be either CCU or CEU as discussed later in the paper) is given by $X_{f_{rx}}=h_k (\sum_{i=1}^{M_f} x_i \sqrt{a_i P_f}) +n_k$, where $n_k$ denotes the channel noise at $k^{th}$ typical FU and $\vert h_k \vert^2$ denotes the total channel gain at $k^{th}$ typical FU.

SINR at $k^{th}$ typical FU to decode message of $j^{th}$ FU ($j<k$) is given by
\begin{equation}{\label{eq:SINRNOMAdecode}}
\text{SINR}_{k\to j}^f=\frac{ \rho_f^r P_f  a_j \vert h_k^f\vert^2}{ \rho_f^r P_f \vert h_k^f\vert^2 \sum _{l=j+1}^{M_f} a_l +\rho_f \mathcal{I}_f+\rho_m \mathcal{I}_m+1},
\end{equation}
where $\rho_f^r=\mathbb{E}[x_i^2]/\sigma_f^2$ denotes the receiving transmit SNR at FU. $a_n$ denotes power allocation factor for FU with index $n={k,j,l}$. $\mathcal{I}_{m}$ denotes the cross-tier interference at typical FU and is given by $\mathcal{I}_{m}=\sum _{x\in \Omega_{m}/\{0\}} \vert h_x \vert ^2$, where $\vert h_{x}\vert^2$ denotes the total channel gain from $x^{th}$ MBS to typical FU assumed to be at origin using Slivnyak's theorem \cite{Slivnyak}. $\mathcal{I}_{f}$ denotes the co-tier interference such that $\mathcal{I}_{f}= \sum _{y\in \Omega_{f}^R/\{0\}} \vert h_y \vert ^2$, where $\vert h_{y}\vert^2$ denotes the total channel gain from $y^{th}$ FBS to typical FU.
SINR at $k^{th}$ typical FU to decode its own message is given by
\begin{equation}{\label{eq:SINRNOMA}}
\text{SINR}_{k}^f=\frac{\rho_f^r P_f  a_k \vert h_k^f\vert^2}{ \rho_f^r P_f \vert h_k^f\vert^2
	\sum _{l=k+1}^{M_f} a_l +\rho_f \mathcal{I}_f+\rho_m \mathcal{I}_m+1}.
\end{equation}

\section{Performance Analysis}  \label{sec:analytical}

This section derives the outage probability of MBS tier and FBS tier with NOMA using carrier sensing. Outage probability of the FBS tier with NOMA includes the outage probability of both type of users i.e., the CCU and the CEU.

\subsubsection{SINR Outage Probability Analysis for MBS Tier}\label{sec:outageMU}
The outage probability of a typical MU is given as follows. 
\begin{prop}\label{prop:OUTMU}
Conditioned on the fact that MU connects to the nearest MBS, the outage probability of a typical MU is given as
\begin{equation}
\mathcal{P}_{O}^m=\pi \lambda_m \mathcal{Y}_m^2 \sum\limits_{n = 0}^{N} {b_{n}^m}e^{-c_{n}^m \frac{\phi} {{\rho_m P_m}} }  e^{\mu_m^f},
\end{equation}
where $N$ is a parameter to ensure a complexity-accuracy trade-off, ${b_{n}^m} = - {w_{N}}\sqrt {1 - \theta _{n}^2} \left({1 \over 2}\left({\theta _{n}+1} \right) \right) e^{-\pi \lambda_m \left({1 \over 2}\left({\theta _{n}+1} \right)\mathcal{Y}_m \right)^2 }$, ${b_0} = - \sum\nolimits_{{n} = 1}^N {b_{n}^m}$, ${c_{n}^m} = 1 + {\left({{{{{ \mathcal{Y}_m}}} \over 2}{\theta _{n}} + {{{{ \mathcal{Y}_m}}} \over 2}} \right)^\nu }$, ${c_0} = 0$, ${w_{N}} = {\pi \over N}$, ${\theta _{n}} = \cos \left({{{2{n} - 1} \over {2{N}}}\pi } \right)$~\cite{random}, and  $\phi=2^{2 R}-1$ denotes SINR threshold. 
\begin{equation}
\mu_m^f=-\lambda_m^R \frac{r_g^{-\alpha' } \left(\alpha_f s_m^f F(r_g,\alpha_f)-(\alpha') K \right)}{\alpha'},
\end{equation}
where $s_m^f=\frac{ c_{n}^m \phi \rho_f} {{ \rho_m P_m }}$, $K=r_g^{\alpha_f} \text{ln} \left(s_m^f r_g^{-\alpha_f
}+1\right)$, $F(r_g,\alpha_f) ={}_2F_1\left(1,\frac{\alpha'}{\alpha_f };2-\frac{1}{\alpha_f};-s_m^f r_g^{-\alpha_f }\right)$ is the hypergeometric function and $\alpha'=\alpha_f-1$.
\end{prop}
	\vspace{1ex}
\textit{Proof}: Please see Appendix \ref{appendix:P1}.\\

\subsubsection{SINR Outage Analysis for FBS Tier (with NOMA)}\label{sec:outageFUNOMA}
The outage probability at the $k^{th}$ typical FU is expressed as
\begin{prop}\label{prop:OUTFUNOMA}
	Conditioned on the uniform distance of a typical FU from FBS and ordered channel gain of the users, the outage probability at the $k^{th}$ typical FU is given as
	
	\begin{multline}\label{eq:outageNOMA}
	P_{k}^f=\psi _k \sum _{z=0}^{{M_f}-k}\left(\begin{array}{c} {M_f}-k \\ z \\ \end{array} \right) \frac{(-1)^z}{k+z} \sum_{T_k^z}  \left(\begin{array}{c} k+z \\ q_0 \ldots q_{N} \\ \end{array} \right) \\
	\left(\prod _{{n}=0}^{N}  {b_{n}^f}^{q_{n}}\right) e^{-\sum _{{n}=0}^N q_{n} c_{n}^f \frac{\epsilon_{max}} {\rho_f  P_f}} \mathcal{L}_{I_m}\left(s_f^m I_m\right) e^{\mu_f^m},
	\end{multline}
	where $\epsilon_{max}=\text{max} \left(\epsilon_1, \epsilon_2, \dots, \epsilon_k \right)$ such that $\epsilon_j$ is evaluated as
	\begin{equation}
	\epsilon_{j}=\frac{\phi_{j}}{\left({{a_{j}} - {\phi _{j}}\sum\limits_{i = {j} + 1}^{M_f} {a_i}} \right)},
	\end{equation}
	where ${\phi _{j}= 2^{R_{j}} - 1}$ and $R_j$ denotes the target data rate of $j^{th}$ user such that $R_j=R ~\forall{j} \in{(1,2,\ldots,{{M_f}})}$. $s_f=\frac{  \epsilon_{max} \sum _{{n}=0}^N q_{n} c_{n}^f}{\rho_f P_f}$, ${b_{n}^f} = - {w_{N}}\sqrt {1 - \theta _{n}^2} \left({{{\mathcal{Y}_f} \over 2}{\theta _{n}} + {{\mathcal{Y}_f} \over 2}} \right)$, ${c_{n}^f} = 1 + {\left({{{\mathcal{Y}_f} \over 2}{\theta _{n}} + {{\mathcal{Y}_f} \over 2}} \right)^\nu }$, ${T_k^z}=\left({{q_0}, \ldots, {q_{N}}}~|~\sum_{i=0}^{N_f} q_i=k+z\right)$, $\psi_k={{{M_f}!} \over {(k - 1)!({M_f} - k)!}}$, $\left(\begin{array}{c} k+z \\ q_0 \ldots q_{N} \\ \end{array} \right) = {{{M_f}!} \over {{q_0}! \ldots {q_{N}}!}}$.

\textbf{Remark 2:} It can be noted that~(\ref{eq:outageNOMA}) contains two expectations terms that contribute to the role of cross-tier and co-tier interference in the outage probability of typical FU. The user with $k=1$ does not perform SIC, hence the term $\epsilon_{max}$ equals $\epsilon_1$. Also the outage probability is dependent on the transmit SNR of both MBS and FBS, and on the user's target rate. The dependence is directly proportional to the target rate and the transmit SNR of MBS tier, while it is inversely proportional to the the transmit SNR of FBS tier as observed in Fig.~\ref{fig:2}.
\end{prop}
\vspace{1ex}
\textit{Proof}: Please see Appendix \ref{appendix:P2}.\\

\subsection{Offloading and NOMA Compatibility (NC) Probability}\label{sec:moreprobabilities}
This section discusses the offloading (OF) probability and the NC probability for the proposed model. OF probability is conditioned on the long term power averaged biased-received-power (BRP) received from the FBS and MBS. The NC probability describes whether the OMU is a CEU or CCU with respect to the available PU at FBS. PU is the FU with which the incoming OMU is paired by the FBS and served using NOMA. Section~\ref{sec:offloading} decides whether MU will be offloaded to FBS tier, and Section~\ref{sec:NC} decides whether OMU is a CCU or CEU with respect to the available PU. 

\subsubsection{OF Probability}\label{sec:offloading}
OF probability from MBS tier to FBS tier can be calculated as follows.
\begin{prop}\label{prop:OFFLOADING}
	Offloading is based on maximum BRP \cite{cellassociationandrews}, where a user is associated with the strongest BS in terms of long-term averaged BRP at the user. The closed form expression for the OF probability for $\nu_m=3$ and $\nu_f=4$ is given as
	
	\begin{multline}
	\mathcal{P}_{}^{m\to f}=-\frac{3}{4} \text{E}\left(\frac{1}{4},\pi  \lambda_m  \left(\frac{B_m P_m}{B_f P_f}\right)^{\frac{1}{2}} \mathcal{Y}_{f}^{8/3}\right)\\
	+\frac{3 \Gamma \left(\frac{3}{4}\right)}{4 ( \pi )^{3/4} \mathcal{Y}_{f}^2 \left(\lambda_m  \left(\frac{B_m P_m}{B_f P_f}\right)^{\frac{1}{2}}\right)^{3/4}}- e^{-\pi  \lambda_m  \mathcal{Y}_{m}^2},
	\end{multline}		
	where $B_m$ and $B_f$ are the bias factor for MBS and FBS tier respectively. $\text{E} (n,x)$ evaluates the exponential integral as $\text{E}(n,x)= \int_1^{\infty } {e^{-xt}}/{t^n} \, dt$ and $\Gamma(x)=\int_0^{\infty } e^{-t} t^{x-1} dt$ is the complete gamma function.
\end{prop}

\textit{Proof}: Please see Appendix \ref{appendix:P3}.\\

\subsubsection{NOMA Compatibility (NC) Probability}\label{sec:NC}			
When the offloaded MU is served using NOMA, it becomes necessary to find out how the OMU will be treated by the FBS, i,e, whether the OMU will be accommodated by the FBS as a CCU or CEU with respect to the available PU. The probability of whether FBS can apply NOMA to the OMU or not is decided on whether the OMU satisfies the sufficiently different channel condition criterion and whether it will be accommodated as a CCU or a CEU. This condition for the OMU is checked with respect to the available PU. Assuming that index $k$ refers to the OMU and $n$ for the available PU, the probability of OMU to be offloaded as a CCU with respect to the PU can be calculated as
\begin{equation}\label{eq:pairingNC}
\mathcal{P}_{NC}=\mathcal{P}\left(\frac{\vert{h_n}{\vert^2} }{\vert{h_k}{\vert^2}} < p\right),
\end{equation}
where $p$ (satisfying $0<p<1$ ) represents the ratio of channel gains PU and OMU. The probability density function (PDF) of the ratio of two order statistics \cite{orderstatpdfratio},\cite{globe} is given as
\begin{multline}\label{eq:ratio}
f_{\frac{h_n^2}{h_k^2}} (z)=\frac{{M_f}!}{{(n-1)! (-n+k-1)! \ ({M_f}-k)!}}\\{ \sum_{j_1=0}^{(n-1)} \sum_{j_2=0}^{(-n+k-1)} \
	\frac{(-1)^{j_1+j_2} \left(
		\begin{array}{c}
		n-1 \\
		j_1 \\
		\end{array}
		\right) \left(
		\begin{array}{c}
		-n+k-1 \\
		j_2 \\
		\end{array}
		\right)}{\left(z \
		t_1+t_2\right){}^2}},
\end{multline}
where $t_1=j_1-j_2+k-n$, and $t_2={M_f}-k+1+j_2$. Hence, the probability can be calculated using $\mathcal{P}_{NC}=\int_{0}^p f_{({h_n^2}/{h_k^2})} (z) dz$. The value of $p$ signifies the amount of difference in the channel gains between the OMU and PU. Hence, we may say that $p$ is a measure of the channel condition of OMU with respect to the available PU. Results for different values of $p$ are discussed in Section~\ref{sec:results}. A lower value of $p$ signifies a large difference in the users' channel gain, while a large value of $p$ signifies smaller difference in the users' channel gain.

\textbf{Remark 3:} A tractable analysis is done with ${M_f}=2$, $k=2$ (OMU), and $n=1$ (PU). Hence, we get the NC probability as $\mathcal{P}_{NC}={2 p}/({p+1})$. NC probability helps us differentiate whether the OMU is a CCU or CEU with respect to the available PU.

\subsection{Total Outage Probability After Offloading}
Combining outage probability, OF probability and NC probability, the total outage probability when a PU is assumed to be available with FBS for the incoming OMU can be written in three cases, depending on the relative channel condition of OMU with respect to the available PU as follows
\begin{itemize}
	\item Case I: When MU is offloaded to FBS (without NOMA)
	\begin{equation}\label{eq:Case1}
	\mathcal{P}_{T}=(1-\mathcal{P}_{}^{m\to f})\mathcal{P}_{O}^m+\mathcal{P}_{}^{m\to f}  \mathcal{P}_{O}^f.
	\end{equation}
	\item Case II: When MU is offloaded as a CCU with respect to available PU at FBS (with NOMA)
	\begin{multline}\label{eq:Case2}
	\mathcal{P}_{{T}}^{C}=(1-\mathcal{P}_{}^{m\to f})\mathcal{P}_{O}^m+\mathcal{P}_{}^{m\to f} \mathcal{P}_{NC} \mathcal{P}_{k}^f.
	\end{multline}
	\item Case III: When MU is offloaded as a CEU with respect to available PU at FBS (with NOMA)
	\begin{multline}\label{eq:Case3}
	\mathcal{P}_{{T}}^{E}=(1-\mathcal{P}_{}^{m\to f})\mathcal{P}_{O}^m+\mathcal{P}_{}^{m\to f} (1-\mathcal{P}_{NC}) \mathcal{P}_{k}^f.
	\end{multline}
\end{itemize}

\textbf{Remark 4:} The above equations in (\ref{eq:Case1}), (\ref{eq:Case2}), and (\ref{eq:Case3}) combine two situations, one where no offloading takes place (denoted by the first terms), and second when offloading occurs (denoted by the second terms). Case II and Case III also includes the NC probability in their second terms as a check for whether the incoming OMU is a CEU or CCU with respect to the available PU. As can be seen from Fig.~\ref{fig:3} and Fig.~\ref{fig:4}, the NC probability effects the performance of the OMU depending on its relative channel condition with respect to the PU.
\begin{table}[h]
	\centering
	\caption{ Network Parameters}
	\label{tab:data}
	\begin{tabular}{|p{2cm}|p{4cm}|} \hline
		\vspace{-1ex}  Symbols & \vspace{-1ex} Value  \\ \hline
		\vspace{-1ex}  $ P_m $, $P_f$ & \vspace{-1ex} 40 W, 1 W \\ \hline
		\vspace{-1ex}  $\lambda_m $  & \vspace{-1ex} $10^{-4}m^{-2}$\\ \hline
		\vspace{-1ex} $\lambda_f $  & \vspace{-1ex} $10^{-3}m^{-2}$ and $10^{-1}m^{-2}$ \\ \hline
		\vspace{-1ex} $a_k$ & \vspace{-1ex} $0.2, 0.8$\\ \hline
		\vspace{-1ex}  $T_B$ & \vspace{-1ex} $0$dB\\ \hline
		\vspace{-1ex}   $\mathcal{Y}_m, \mathcal{Y}_f$ & \vspace{-1ex}  $1 \text{km}, 5 \text{m}$ \\ \hline

	\end{tabular}
\end{table}

\section{Results and Discussions}\label{sec:results}
\begin{figure}[h]
\center
	\includegraphics[height=7cm,width=9cm]{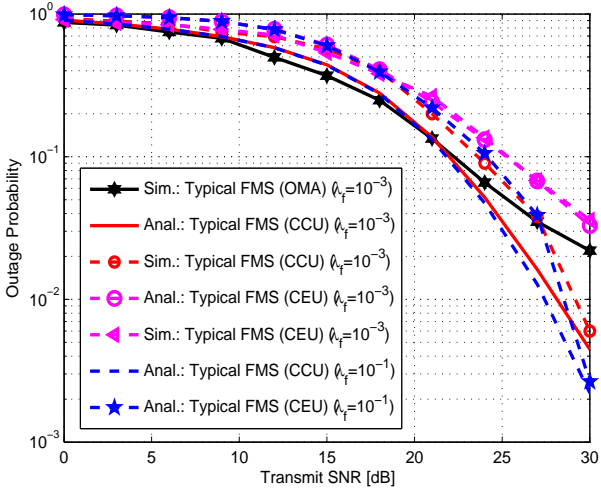}\\
	\caption{ Variation of outage probability with transmit SNR for different FBS density ($\lambda_m=10^{-4}$). }\label{fig:2}
\end{figure}
 \begin{figure}
	\center
	\includegraphics[height=7cm,width=9cm]{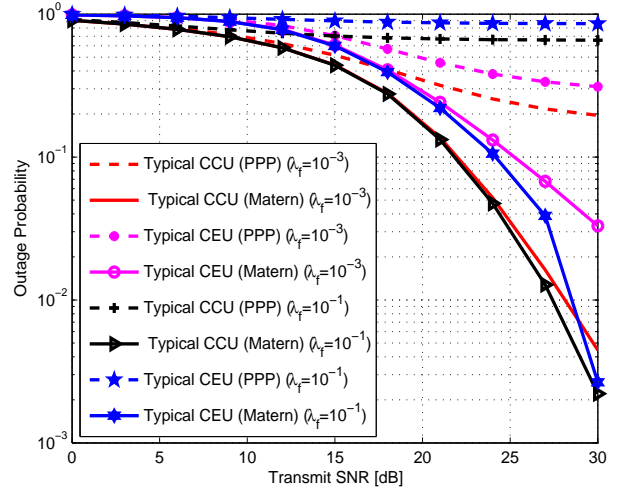}\\
	\caption{Comparison of outage probability for PPP and RPP modeling at different FBS density ($\lambda_m=10^{-4}$). }\label{fig:3}
\end{figure}
In this section, outage probability of the NOMA-HetNet with carrier sensing is studied based on the analytical expression derived in Section~\ref{sec:analytical} for the two-tier HetNet, where FBS tier uses NOMA and carrier sensing for its transmissions. The transmit SNR is varied from 0 to 30 dB for both the tiers and $N=10$. Transmit SNR at MBS tier and FBS tier is considered to be fixed at $\rho_m=16dB$ and $\rho_f=0dB$ \cite{reference}, while analyzing the FBS tier and MBS tier performance, respectively. The graphs shows analytical (Anal.) curves verified using Monte Carlo simulation (Sim.) curves.

Fig.~\ref{fig:2} shows the variation of outage probability with transmit SNR of the FBS tier for different FBS densities. Also, for comparative study, outage probability of FBS tier using OMA (referred as FBS-OMA) modeled with the same carrier sensing, as used for FBS-NOMA, has been plotted. From simulations it is observed that for low transmit SNRs the performance of both FBS-NOMA and FBS-OMA using carrier sensing are nearly same however at higher transmit SNR, FBS-NOMA surpasses the performance of FBS-OMA. The performance of FBS-OMA degrades due to increase in co-tier interference at high transmit SNR from interferers in the vicinity. The improvement shown by FBS-NOMA using RPP results in decrease in the outage probability by $78.57 \%$ as compared to FBS-OMA with RPP. Construction of RPP (to enable the proposed carrier sensing in FBS tier) from parent PPP removes the FBS that do not fulfill the hard core parameter criterion (or minimum distance criterion between FBSs). This leads to the removal of nearby interferers, that have a large contribution in the total interference at typical FU. Clearly, for a dense FBS network the number of such removals will be higher as compared to a sparse FBS network. Hence, for a higher density the number of FBSs removed will be more as compared to when the FBS density is assumed to be low. This renders a major impact on the net interference at typical FU and hence also on the outage performance of FBS-NOMA. Since, the dominant interferers are removed, the outage probability of a FU (both CEU and CCU) decreases. It is worth pointing that the increase in FBS density (from $10^{-3}$ to $10^{-1}$) has a higher impact on CEU ($90.30 \%$ decrease in outage probability) as compared to CCU ($52.10 \%$ decrease in outage probability) as can also be observed from Fig.~\ref{fig:2}. Since a CCU is already present near to an FBS, increasing FBS density does not impact the CCU much. However, as mentioned earlier, CEU is farther away from FBS and has poorer channel condition as compared to the CCU. One way to improve the quality of service of CEU is to increase the density of FBS such that chances of an FBS lying close to CEU increases. However, higher density also implies higher co-tier interference hence, increasing the density does not always imply an improved performance. Employing carrier sensing on FBS-NOMA and hence using RPP to model the FBS-NOMA network with higher density, instead of using PPP, guarantees an increased chances of an FBS lying closer to CEU in addition to managed co-tier interference due to the thinning process from carrier sensing. Hence, increasing the density of FBS tier leads to larger decrease in outage probability for a CEU as compared to CCU. This improvement is suppressed in high density FBS network modeled using PPP due to increased co-tier interference from large number of FBS as shown in Fig.~\ref{fig:3}. Hence, we may infer that RPP also caters to the well known issue of performance enhancement in terms of decreased outage probability for CEUs by compensating the drawback of increased co-tier interference of PPP modeled FBS tier at higher densities.

\begin{figure*}[t]
\hfill
\begin{subfigure}[For $p$=0.1.]{\includegraphics[width=5.8 cm,height=5cm]{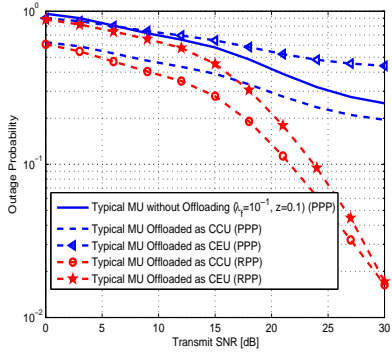}}
\label{fig:4zeq0pt1}
\end{subfigure}
\hfill
\begin{subfigure}[For $p$=0.5.]{\includegraphics[width=5.8 cm,height=5cm]{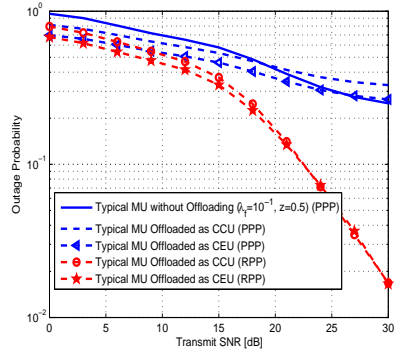}}
\label{fig:4zeq0pt5}
\end{subfigure}
\hfill
\begin{subfigure}[For ${p}$=0.8.]{\includegraphics[width=5.8 cm,height=5cm]{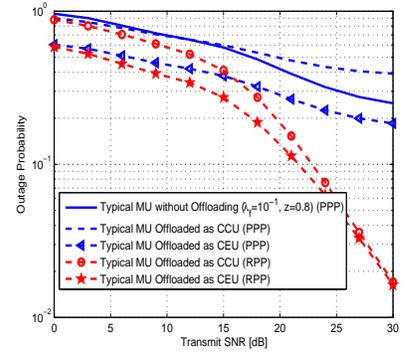}}
\label{fig:4zeq0pt8}
\end{subfigure}
\hfill
\caption{Variation of outage probabilities with transmit SNR for different value of $p$ ($\lambda_f=10^{-1}$).}
\label{fig:5}
\end{figure*}

Fig.~\ref{fig:3} shows the comparison of outage probability of FBS-NOMA tier for the two cases, when the network is modeled using PPP and using RPP. The current literature shows the performance enhancement of NOMA over OMA, however uses PPP assumptions for modeling the BSs. When the NOMA network is modeled using RPP, it gives even better results as compared to when the network is modeled using PPP as observed from Fig.~\ref{fig:3}. The reason, as discussed earlier, is the reduced co-tier interference due to the removal of dominant interferers, due to the thinning  process of RPP, that otherwise hinders the performance of high density networks. As the density of FBS tier is increased for both PPP and RPP model, it is observed that higher FBS density increases the outage probability of FBS-NOMA modeled using PPP while decreases the outage probability of FBS-NOMA modeled using RPP. Carrier sensing manages the interference and the increasing density has a positive impact on the FBS tier performance instead of a negative effect as seen for PPP. As an observation it can also be noticed that the performance improvement (between PPP and RPP) for CCU ($74.04 \%$ decrease in outage probability) is higher for low density as compared to CEU ($ 48.65 \%$ decrease in outage probability). However, for higher density this performance enhancement becomes nearly the same ($ 99.6 \%$ decrease in outage probability) for both CCU and CEU. Hence, we may conclude that FBS-NOMA network modeled using RPP gives better performance, especially for CEU, as compared to PPP modeling.
\begin{figure*}[t]
\hfill
\begin{subfigure}[For $p$=0.1.]{\includegraphics[width=5.8 cm,height=5cm]{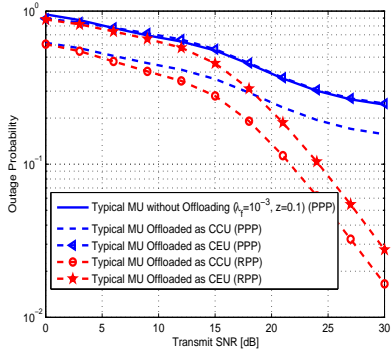}}
\label{fig:3zeq0pt1}
\end{subfigure}
\hfill
\begin{subfigure}[For $p$=0.5.]{\includegraphics[width=5.8 cm,height=5cm]{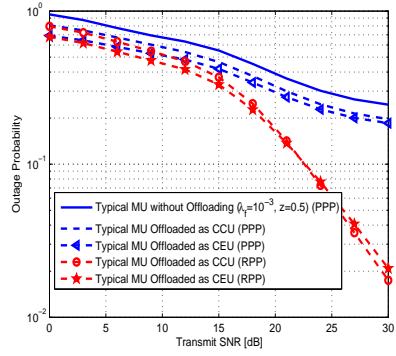}}
\label{fig:3zeq0pt5}
\end{subfigure}
\hfill
\begin{subfigure}[For ${p}$=0.8.]{\includegraphics[width=5.8 cm,height=5cm]{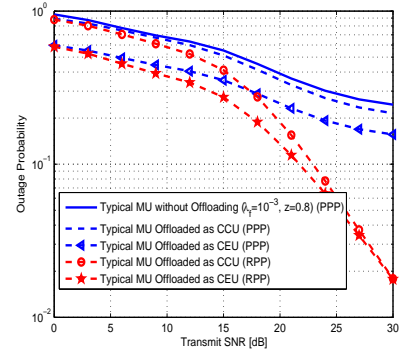}}
\label{fig:3zeq0pt8}
\end{subfigure}
\hfill
\caption{Variation of outage probabilities with transmit SNR for different value of $p$ ($\lambda_f=10^{-3}$).}
\label{fig:4}
\end{figure*}
Fig.~\ref{fig:4} shows the total outage probability of a typical MU after offloading to FBS tier for an FBS density of $\lambda_f=10^{-3}$. The figures are plotted for three different values of $p$, signifying the different channel conditions of the MU at the time of offloading. A comparison for the two cases, i.e., one where no carrier sensing on FBS-NOMA tier is used and other where carrier sensing is incorporated in the FBS-NOMA tier, is done. Fig.~\ref{fig:4} and Fig.~\ref{fig:5} also shows the outage probability of typical MU when offloading is not performed and is compared with the offloading scenario. As can be seen from Fig.~\ref{fig:4} and Fig.~\ref{fig:5}, offloading to the FBS-NOMA tier modeled using RPP yields a better outage probability, in all the three cases, and for both CCU and CEU, when compared to the offloading to FBS-NOMA tier modeled using PPP assumptions. Also, when compared with the outage probability of typical MU without offloading, we may observe that offloading to FBS-NOMA modeled using RPP yields better outage probability which is not always the case for offloading to FBS-NOMA modeled using PPP.

From Fig.~\ref{fig:4} it can be observed that for the PPP modeling, when $p=0.1$, i.e, when the difference in OMU and PU is large, offloading as a CCU yields a decreased outage probability while offloading as a CEU does not give any visible improvement in the outage probability. This situation is reversed when $p=0.8$, i.e., when the difference in channel gain between OMU and PU decreases. For $p=0.8$, CEU performance is enhanced while the offloaded CCU does not show any improvement as compared to no offloading. For $p=0.5$, the difference in the channel gain between OMU and PU is larger than that of $p=0.8$ and smaller than $p=0.1$. Hence, for $p=0.5$, the OMU offloaded either as CCU or CEU yields nearly same outage performance.
Next, we increase the density of FBS tier from $\lambda_f=10^{-3}$ to $\lambda_f=10^{-1}$. Fig.~\ref{fig:5} shows the total outage probability of the OMU as a CCU or CEU to FBS-NOMA tier for an FBS density of $\lambda_f=10^{-1}$. Again, similar to Fig.~\ref{fig:4}, the graphs are plotted for different channel condition of the OMU. It is observed that with the increased FBS density, the impact of carrier sensing can be seen more adequately. It can be noted from Fig.~\ref{fig:5} that for higher FBS density the offloading to FBS-NOMA tier without carrier sensing degrades the performance of the OMU for some cases. However, with carrier sensing at FBS-NOMA a gain in outage performance is seen for all the three cases of offloading. This is because without carrier sensing at FBS-NOMA, increased density of FBS also increases the aggregate interference at the OMU, while the increased interference is managed by using carrier sensing. Fig.~\ref{fig:5} (a) is plotted for a value of $p=0.1$ which implies that the difference in channel condition between the OMU and PU is large. This implies that for $p=0.1$ the offloaded CEU will have a much poorer channel condition and offloaded CCU will have a much better channel condition as compared to its corresponding PU. As can be seen from the curves, when modeling of FBS-NOMA is done using PPP assumption, due to lack of interference management the offloaded CEU's performance is degraded as compared to when no offloading is done. However, a good channel condition for the offloaded CCU decreases the outage probability for the OMU. Similarly, Fig.~\ref{fig:5} (c) is plotted for $p=0.8$ which implies that there is not much difference in channel condition between OMU and PU. This indicates that the channel condition of OMU as CCU is not as good as compared to when $p=0.1$. This leads to degradation of OMU's performance when offloaded as a CCU. This is because the power allocation factors are fixed for CEU and CCU to be $0.2$ and $0.8$, respectively, and therefore even though CCU's channel condition is not as good for $p=0.8$ as it was for $p=0.1$, it is served by the same power as for $p=0.1$. This leads to the increase in outage probability of OMU as CCU. The interference management at the FBS-NOMA tier using RPP compensates any such degradation seen at offloaded CEU or CCU. Thus, we may conclude, no carrier sensing at FBS-NOMA tier leads to the degradation in outage performance of either the offloaded CEU or offloaded CCU, depending on their channel condition during offloading. However, the use of carrier sensing (or modeling using RPP), positions the active FBS such that the interference at the OMU is managed, and thus unlike PPP, none of the three cases of offloading results in degradation at OMU. Hence, carrier sensing in FBS-NOMA tier plays a crucial role in the interference management and hence in the performance gain at OMU from offloading.

\section{Conclusion}\label{sec:conclusion}

This work presents a mathematical framework of HetNet comprising MBS tier and FBS tier. The FBS tier uses NOMA and carrier sensing for transmission. The carrier sensing is modeled using an RPP. Offloading of MU from MBS tier to FBS tier helps in load balancing in HetNets. The offloading is studied under different channel conditions of OMU with respect to available PU at FBS tier and some useful observations are drawn. The comparison of the proposed carrier sensing model in FBS-NOMA tier is done with two existing techniques namely, FBS-NOMA without carrier sensing and FBS-OMA. Both the comparisons supports the superiority of the proposed model. It is also observed that the use of carrier sensing in high density FBS-NOMA tier (modeled using RPP) provides decreased outage probability for OMU in all the channel conditions during offloading unlike when the FBS-NOMA tier did not performed carrier sensing (modeled using PPP). Thus, the RPP model and its analysis of NOMA-HetNet is vital for 5G and beyond communication systems.

\appendices
\section{Proof of Proposition 1}\label{appendix:P1}

Assume that a typical MU connects to the nearest MBS, small scale fading is Rayleigh distributed, and users follows homogeneous PPP distribution, then by applying the polar coordinates, the cumulative density function (CDF) of the unordered channel gain of MBS tier can be written as \cite{random}
\begin{equation}\label{eq:CDFMBS}
	{F_{\vert\tilde{h}_m{\vert^2}}}(y) = 2 \pi \lambda_m \int_0^{{{ \mathcal{Y}_m}}} \left({1 - {e^{- (1 + {r_m^{\nu_m} })y}}} \right) e^{-2 \pi \lambda_m r_m^2} r_m dr_m.
\end{equation}
Using G-C quadrature \cite{GCQuad}, (\ref{eq:CDFMBS}) can be approximated as
\begin{equation}\label{eq:CDFMBS2}
	{F_{\vert\tilde{h}_m{\vert^2}}}(y) \approx \pi \lambda_m \mathcal{Y}_m^2 \sum\limits_{n = 0}^N {b_{n}^m}e^{-c_{n}^m y},
\end{equation}	
where $N$ is a parameter to ensure a complexity-accuracy trade-off, ${b_{n}^m} = - {w_{N}}\sqrt {1 - \theta _{n}^2} \left({1 \over 2}\left({\theta _{n}+1} \right) \right) e^{-\pi \lambda_m \left({1 \over 2}\left({\theta _{n}+1} \right)\mathcal{Y}_m \right)^2 }$, ${b_0} = - \sum\nolimits_{{n} = 1}^N {b_{n}^m}$, ${c_{n}^m} = 1 + {\left({{{{{ \mathcal{Y}_m}}} \over 2}{\theta _{n}} + {{{{ \mathcal{Y}_m}}} \over 2}} \right)^{\nu_m} }$, $
{c_0} = 0$, ${w_{N}} = {\pi \over N}$, ${\theta _{n}} = \cos \left({{{2{n} - 1} \over {2{N}}}\pi } \right)$.
The outage probability at a typical MU is given as following
\begin{align}\label{eq:MUoutagesteps}
\mathcal{P}_{O}^{m}&=\mathcal{P} \left(\alpha_m\times \text{log}(1+\text{SINR}_{{m}})<R \right),\\
&= \mathcal{P} \left({ \vert \tilde{h}_m \vert^2  }<\frac{\phi} {{\rho_m P_m}} \left({1+\rho_f I_f}\right)\right),\nonumber\\
&={F_{\vert\tilde{h}_m{\vert^2}}}\left(\frac{\phi} {{\rho_m P_m}} \left({1 +\rho_f I_f}\right)\right), \nonumber\\
&\stackrel{\text{(a)}}=\pi \lambda_m \mathcal{Y}_m^2 \sum\limits_{n = 0}^N {b_n^m} e^{-c_n^m \frac{\phi} {{\rho_m P_m}} \left({1 +\rho_f I_f}\right)},\nonumber\\
&=\pi \lambda_m \mathcal{Y}_m^2 \sum\limits_{n = 0}^N {b_n^m} e^{- \frac{c_n^m \phi} {{\rho_m P_m}}}\mathbb{E}_{I_f}\left[e^{- \frac{ c_n^m \phi \rho_f I_f} {{\rho_m P_m}}}\right],\nonumber\\
&=\pi \lambda_m \mathcal{Y}_m^2 \sum\limits_{n = 0}^N {b_n^m} e^{- \frac{c_n^m \phi} {{\rho_m P_m}}} \times \mathcal{F}_I,\nonumber\\
\end{align}
where (a) follows from (\ref{eq:CDFMBS2}) and $\alpha_m$ is the fraction of bandwidth allocated to typical MU, $\mathcal{F}_I=\mathbb{E}_{I_f}\left[e^{- s_m^f I_f}\right]$, $s_m^f=\frac{ c_n^m \phi \rho_f} {{\rho_m P_m}}$ and $\phi=2^{2 R}-1$ denotes SINR threshold.
Now, we calculate the cross-tier interference at typical MU from FBS tier as follows similar to \cite{basemaths1}.
\begin{align}
\mathbb{E}_{\mathcal{I}_{{f}}}\left[e^{-s_m^f \mathcal{I}_{{f}} }\right]&=\mathbb{E}_{\mathcal{I}_{{f}}}\left[e^{-s_m^f \sum _{v\in \Omega_{f}^R/\{0\}} \vert h_v \vert ^2 }\right],\\\nonumber
&\stackrel{\text{(a)}}=\mathbb{E}_{\mathcal{I}_{{f}}}\left[e^{-s_m^f \sum _{v\in \Omega_{f}^R/\{0\}} \vert \hat{h}_v \vert ^2 r_v^{-\alpha_f} }\right],\\\nonumber
&=\mathbb{E}_{\Omega_{f}^R/\{0\}}\left[\prod_{v \in \Omega_{f}^R/\{0\}} \mathbb{E}_{\hat{h}_v} \left[ e^{-s_m^f \vert \hat{h}_v \vert ^2 r_v^{-\alpha_f} }\right]\right],\\\nonumber
&=\mathbb{E}_{\Omega_{f}^R/\{0\}}\left[ e^{-\sum_{\Omega_{f}^R/\{0\}} \text{ln}\left( 1+s_m^f r_v^{-\alpha_f}\right) } \right],\\\nonumber
&\stackrel{\text{(b)}}\ge e^{\mathbb{E}_{\Omega_{f}^R/\{0\}}\left[ {-\sum_{\Omega_{f}^R/\{0\}} \text{ln}\left( 1+s_m^f r_v^{-\alpha_f}\right) } \right]},  \\\nonumber
\end{align}
where (a) follows from the assumption of a guard zone around receivers $r_g>1$, hence, the bounded path loss model is reduced to simply $r_v^{-\alpha_f}$ for the calculation of interference, where $r_v$ is the distance between $v^{th}$ FBS to typical MU and (b) follows from Jensen's inequality. 
Let, $\mu_m^f={\mathbb{E}}_{\Omega_{f}^R/\{0\}}\left[-\sum_{v\in\Omega_{f}^R/\{0\}}\Delta_{v}\right],$
where $\Delta_{v}=\text{ln}\left( 1+s_m^f r_v^{-\alpha_f}\right) $. Using Campbell's theorem \cite{dhillon}, we can write as
\begin{align}
\label{eq:mu} \mu_m^f&={\mathbb{E}}_{\Omega_{f}^R}^{!o}\left[-\sum_{v\in\Omega_{f}^R}\Delta_{v}\right]=\int_{r_g}^{\infty} \lambda_f^R \Delta_{v}(r_v) d r_v,\\\label{eq:mu}\nonumber
&=\int_{r_g}^{\infty} - \lambda_f^R \text{ln}\left( 1+s_m^f r_v^{-\alpha_f}\right)  d r_v,\\ \nonumber
&=-\lambda_f^R \frac{r_g^{-\alpha' } \left(\alpha_f s_m^f F(r_g,\alpha_f)-(\alpha') r_g^{\alpha_f} \text{ln} \left(s_m^R r_g^{-\alpha_f
	}+1\right)\right)}{\alpha'},
\end{align}
where $F(r_g,\alpha_f) ={}_2F_1\left(1,\frac{\alpha'}{\alpha_f };2-\frac{1}{\alpha_f};-s_m r_g^{-\alpha_f }\right)$ is the hypergeometric function and $\alpha'=\alpha_f-1$. 

\section{Proof of Proposition 2}\label{appendix:P2}

Using the assumption of homogeneous PPP, the CDF of unordered channel gain of FU can be expressed as \cite{random},
\begin{equation}\label{eq:CDFFBS}
{F_{\vert\tilde{h}_f{\vert^2}}}(y) = {2 \over {\mathcal{Y}_f^2}}\int_0^{\mathcal{Y}_f} \left({1 - {e^{- (1 + {z^{\alpha_f} })y}}} \right)z\,dz.
\end{equation}
By applying the G-C quadrature \cite{GCQuad} to (\ref{eq:CDFFBS}), we get
\begin{equation}\label{eq:CDFFBS2}
{F_{\vert\tilde{h}_f{\vert^2}}}(y) \approx {1 \over {\mathcal{Y}_f}}\sum\limits_{n = 0}^{N} {b_{n}^f}e^{-c_{n}^f y},
\end{equation}
where  ${b_{n}^f} =-{w_{N}}\sqrt {1 - \theta _{n}^2} \left({{{\mathcal{Y}_f} \over 2}{\theta _{n}} + {{\mathcal{Y}_f} \over 2}} \right)$, ${c_{n}^f} = 1 + {\left({{{\mathcal{Y}_f} \over 2}{\theta _{n}} + {{\mathcal{Y}_f} \over 2}} \right)^{\alpha_f} }$.

The  ordered channel gain of FBS tier is related with the unordered channel gain of FBS tier $F_{\vert{\tilde{h}_f}{\vert^2}}(y)$ \cite{underlay} as
\begin{equation}\label{eq:CDFFBS3}
{F_{\vert{h_k^f}{\vert^2}}}(y)=\psi _k \sum _{z=0}^{{M_f}-k}\left(\begin{array}{c} {M_f}-k \\ z \\ \end{array} \right) \frac{(-1)^z}{k+z} \left( {F_{\vert{\tilde{h}_f}{\vert^2}}}(y)\right)^{z+k},
\end{equation}
where $\psi_k={{{M_f}!} \over {(k - 1)!({M_f} - k)!}}$. Substituting (\ref{eq:CDFFBS2}) in (\ref{eq:CDFFBS3}) and applying multinomial theorem we get the CDF of ordered channel gain as
\begin{multline}\label{eq:CDFFBS4}
{F_{\vert{h_k^f}{\vert^2}}}(y)=\psi _k \sum _{z=0}^{{M_f}-k}\left(\begin{array}{c} {M_f}-k \\ z \\ \end{array} \right) \frac{(-1)^z}{k+z} \sum_{T_k^z}  \left(\begin{array}{c} k+z \\ q_0 \ldots q_{N} \\ \end{array} \right) \\
\left(\prod _{{n}=0}^{N}  {b_{n}^f}^{q_{n}}\right) e^{-\sum _{{n}=0}^N q_{n} c_{n}^f y},
\end{multline}
where ${T_k^z}=\left({{q_0}! \ldots {q_{N}}!}~|~\sum_{i=0}^{N} q_i=k+z\right)$, $\left(\begin{array}{c} k+z \\ q_0 \ldots q_{N} \\ \end{array} \right) = {{{M_f}!} \over {{q_0}! \ldots {q_{N}}!}}$. 

Assuming the channel gains of ${M_f}$ users to be ordered as $\vert{h_1^f}{\vert^2} \leq \ldots \leq \vert{h_{M_f}^f}{\vert^2}$, and hence the corresponding power allocation factors as ${a_1} \geq \ldots \geq {a_{M_f}}$ we derive the outage probability at $k^{th}$ FU as

\begin{equation}
P_{k}^f = \mathcal{P}\left(\text{SINR}_{k \to j}^f < {\phi _j}, \text{SINR}_{k}^f < {\phi _k}\right),
\end{equation}
where $\phi_n=2^{R} - 1$ such that $n$ denotes user index, $\text{SINR}_{k \to j}$ and $\text{SINR}_{k}$ are given in (\ref{eq:SINRNOMAdecode}) and (\ref{eq:SINRNOMA}), respectively.

We observe that the first user (i.e., $k=1$), according to the ordered channel gains, does not perform any SIC. All users after it (i.e., $k>1$) decodes the information of users preceding them  (i.e., $j<k$), and then decode their own message. Since the outage probability is decided based on successful SIC followed by successful decoding of self message, we can write outage probability of $k^{th}$ user as
\begin{equation}
P_{k}^f = \mathcal{P}\left({\vert{h_k^f}{\vert^2} < \frac{\epsilon_{max} (1+\rho_f \mathcal{I}_f+\rho_m \mathcal{I}_m)}{\rho_f P_f  }}\right),
\end{equation}
where $\epsilon_{max}=\text{max} \left(\epsilon_1, \epsilon_2, \dots, \epsilon_k \right)$ such that $\epsilon_j$ is calculated as
\begin{equation}
\epsilon_{j}=\frac{\phi_{j}}{\left({{a_{j}} - {\phi _{j}}\sum\limits_{i = {j} + 1}^{M_f} {a_i}} \right)},
\end{equation}
where ${\phi _{j}= 2^{R_{j}} - 1}$ and $R_j$ denotes the target data rate of $j^{th}$ user such that $R_j=R ~\forall{j} \in{(1,2,\ldots,{{M_f}})}$. This gives the outage probability as
\begin{equation}\label{eq:out}
P_k^f=F_{\vert h_k^f \vert^2}(y),
\end{equation}
where $y=\frac{\epsilon_{max} (1+\rho_f \mathcal{I}_f+\rho_m \mathcal{I}_m)}{\rho_f P_f  }$. Hence, the outage probability of  $k^{th}$ user can be calculated using (\ref{eq:out}) and (\ref{eq:CDFFBS4}) as

\begin{multline}
P_{k}^f=\psi _k \sum _{z=0}^{{M_f}-k}\left(\begin{array}{c} {{M_f}}-k \\ z \\ \end{array} \right) \frac{(-1)^z}{k+z} \sum_{T_k^z}  \left(\begin{array}{c} k+z \\ q_0 \ldots q_{N} \\ \end{array} \right) \\
\left(\prod _{{n}=0}^{N}  {b_{n}^f}^{q_{n}}\right) e^{-\sum _{{n}=0}^N q_{n} c_{n}^f \frac{\epsilon_{max}} {\rho_f  P_f}} \mathbb{E}_{\mathcal{I}_{{m}}}\left[e^{-s_m \mathcal{I}_{{m}} }\right] \times \\ \mathbb{E}_{\mathcal{I}_{{f}}}\left[e^{-s_f \mathcal{I}_{{f}}}\right],
\end{multline}
where $s_f^m=   \frac{ \rho_{m} \epsilon_{max} \sum _{s=0}^N q_n c_n^f}{\rho_f P_f}$ and $s_f^f= \sum _{{n}=0}^N q_{n} c_{n}^f \frac{\epsilon_{max}} { P_f} $.
Hence, we write the outage probability as
\begin{multline}
P_{k}^f=\psi _k \sum _{z=0}^{{M_f}-k}\left(\begin{array}{c} {{M_f}}-k \\ z \\ \end{array} \right) \frac{(-1)^z}{k+z} \sum_{T_k^z}  \left(\begin{array}{c} k+z \\ q_0 \ldots q_{N} \\ \end{array} \right) \\
\left(\prod _{{n}=0}^{N}  {b_{n}^f}^{q_{n}}\right) e^{-\sum _{{n}=0}^N q_{n} c_{n}^f \frac{\epsilon_{max}} {\rho_f  P_f}} \mathcal{L}_{I_m}(s_f^m) \times e^{\mu_f^f},
\end{multline}

Laplace transform of cross tier interference from MBS tier and is calculated as
\begin{equation}
\mathcal{L}_{\mathcal{I}_{{m}}}(s)=e^{\pi \lambda_m  \left(s^{\delta_m } \Gamma(1-\delta_m , s) -s^{\delta_m } \Gamma(1-\delta_m) \right)},
\end{equation}
where $\delta_m=2/\nu_m$, $\Gamma(a,x)=\int_{x}^{\infty} t^{a-1} e^{-t}$ and $\Gamma(z)=\int_{0}^{\infty} x^{z-1} e^{-x}$. 
For co-tier interference, the interference is considered beyond the tagged BS. Hence, we replace $r_g$ by $\mathcal{Y}_f$ in (\ref{eq:mu}) to calculate the co-tier interference.

\section{Proof of Proposition 3}\label{appendix:P3}
Offloading is based on maximum BRP \cite{cellassociationandrews} and a user is associated with the strongest BS in terms of long-term averaged BRP at the user. Hence, the offloading probability can be calculated as follows
\begin{align}
\mathcal{P}_{}^{m\to f}&=\mathbb{E}_{r_f}\left[\mathcal{P} \left(B_m P_m r_m^{-\nu_m }<B_f P_f r_f^{-\nu_f }\right)\right],\\
&=\mathbb{E}_{r_f}\left[\left(e^{- \pi  \lambda_m  r_f^{\frac{2 \nu_f}{\nu_m}}
	\left(\frac{B_m P_m}{B_f P_f}\right)^{\frac{2}{\nu_m}} }-e^{- \pi  \lambda_m
	\mathcal{Y}_m^2}\right)\right], \nonumber
\end{align}%
where $B_m$ and $B_f$ are the bias factor for MBS and FBS tier respectively. The probability distribution of $r_f$ can be expressed as $f(r_f)=2 r_f/\mathcal{Y}_{f}^2$ assuming uniform distribution of FU around FBS within
radius $\mathcal{Y}_f$ and $r_m$ follows $f(r_m)=2 \pi r_m \lambda_{m} \times e^{-\pi r_m^2 \lambda_{m}}$ owing to NN policy. The path loss exponent is taken as $\nu_m=3$ for MBS tier and $\nu_f=4$ for FBS tier. Using these values we get a closed form expression for the offloading probability as

\begin{multline}
	\mathcal{P}_{}^{m\to f}=-\frac{3}{4} \text{E}\left(\frac{1}{4},\pi  \lambda_m  \left(\frac{B_m P_m}{B_f P_f}\right)^{\frac{1}{2}} \mathcal{Y}_{f}^{8/3}\right)\\
	+\frac{3 \Gamma \left(\frac{3}{4}\right)}{4 ( \pi )^{3/4} \mathcal{Y}_{f}^2 \left(\lambda_m  \left(\frac{B_m P_m}{B_f P_f}\right)^{\frac{1}{2}}\right)^{3/4}}- e^{-\pi  \lambda_m  \mathcal{Y}_{m}^2},
\end{multline}		
where $\text{E} (n,x)$ evaluates the exponential integral as $\text{E}(n,x)= \int_1^{\infty } {e^{-xt}}/{t^n} \, dt$ and $\Gamma(x)=\int_0^{\infty } e^{-t} t^{x-1} dt$ is the complete gamma function.

\begin{spacing}{0.92}
\bibliographystyle{IEEEtran}
\bibliography{refJ2}
\end{spacing}
\end{document}